\documentclass[11pt]{article}
\usepackage{latexsym}

\usepackage{amsmath}

\usepackage{amsfonts}

\usepackage{amssymb}

\newtheorem{thm}{Theorem}[section]

\newtheorem{lem}[thm]{Lemma}

\newcommand{\ms}{\medskip}

\newcommand{\ra}{\rightarrow}
\newcommand{\bea}{\begin{eqnarray}}
\newcommand{\eea}{\end{eqnarray}}

\newcommand{\q}{{\cal Q}}

\newcommand{\h}{{\mathcal H}}
\newcommand{\K}{{\mathcal K}}

\newcommand{\fd}{finite-dimensional}
\newcommand{\id}{infinite-dimensional}
\newcommand{\fcd}{finite-codimensional}
\newcommand{\pa}{Poisson algebra}
\newcommand{\pb}{Poisson bracket}
\newcommand{\la}{Lie algebra}

\renewcommand{\ss}{semisimple}

\newcommand{\fb}{{\mathfrak b}}

\newcommand{\fc}{{\mathfrak c}}

\newcommand{\A}{{\mathcal A}}

\def\endproof{\hfill $\blacksquare$}

\newcommand{\C}{{\mathbb C}}
\def\r{{\mathbb R}}
\def\z{{\mathbb Z}}

\begin{document}

\title{On quantizing semisimple basic algebras, II: \\ The general case}

\author{{\bf Mark J. Gotay}\thanks{Supported in part by NSF grant
DMS  00-72434.}  \\
Department of Mathematics \\ University of Hawai`i \\ 2565 The
Mall \\ Honolulu, HI 96822  USA \\ {\footnotesize E-mail:
gotay@math.hawaii.edu}
}
\date{November 30, 2001 \\(Revised \today)}
\maketitle
\normalsize
\vspace{-0.1in}

\begin{abstract}

We prove that there is no consistent polynomial quantization of the
coordinate ring of a basic non-nilpotent coadjoint orbit of a semisimple
Lie group. 
\end{abstract}

\clearpage

\newpage


\section{Introduction}\label{intro}

In a recent paper (Gotay [2002]) we showed that there do not exist
polynomial quantizations of the coordinate ring $P(M)$ of a basic
semisimple coadjoint orbit $M \subset {\rm sl}(2,\r)^*$. Here we extend
this result to any basic non-nilpotent coadjoint orbit of a general \ss\
Lie group:

\begin{thm}
Let $\fb$  be a \fd\ \ss\ \la, and $M$ a
basic non-nilpotent
coadjoint orbit in $\fb^*$. Then
there are no polynomial quantizations of the coordinate ring $P(M)$.
\label{thm:nogo}
\end{thm}

We refer the reader to Gotay [2000] for definitions and
discussions of basic coadjoint orbits and quantization. 

In particular, a coadjoint orbit $M \subset \fb^*$ is \emph{basic}
provided
$\fb$ has no subalgebras which act transitively on $M$ and which globally
separate points of $M$.
Unfortunately, it is difficult to determine exactly which
orbits are basic. From
\S4 in Gotay, Grabowski, and Grundling [2000] and Prop.~2.1 in Gotay
[2002] we know that $M \subset \fb^*$ will be basic for a \ss\ \la\ $\fb$
whenever

\begin{enumerate}
\item[](\emph{i}) $\fb$ is compact and $M$ is principal,
\item[](\emph{ii}) $\fb$ is compact and simple, and
\item[](\emph{iii}) $M$ is principal nilpotent.
\end{enumerate}

Consider the symmetric algebra
$S(\fb)$, regarded as the ring of polynomials on $\fb^*.$ The Lie
bracket on $\fb$ may be extended via the Leibniz rule to a Poisson
bracket on $S(\fb)$, so that the latter becomes a Poisson algebra. Let
$I(M)$ be the associative ideal in $S(\fb)$ consisting of all
polynomials which vanish on $M$ and set $P(M) = S(\fb)/I(M)$. Since
$M$ is an orbit $I(M)$ is also a Lie ideal, hence a Poisson ideal, so the
coordinate ring $P(M)$ of $M$ inherits the structure of a \pa\ from
$S(\fb)$. We denote the \pb s on both $P(M)$ and $S(\fb)$ by
$\{\cdot,\cdot\}$.

Here we are interested in quantizing the coordinate ring
$P(M)$. By a \emph{quantization} of $P(M)$  we mean a Lie representation
$\q$ thereof by symmetric operators preserving a fixed dense domain $D$ in
some separable Hilbert space $\h$, such that $\q
\!\restriction\! \fb$ is irreducible, integrable, and faithful.  Let
$\A$ be the associative operator algebra generated over $\C$ by $I$ and 
$\{\q(b)
\mid b
\in
\fb\}$. We say that a quantization $\q$ of $P(M)$ is \emph{polynomial} if
$\q$ is valued in $\A$.


\section{Proof of Theorem \ref{thm:nogo}}
\label{sec:pf}

Suppose to the contrary that $\q$ were a
polynomial quantization of 
$P(M)$ in a dense invariant domain $D$ in a Hilbert space $\h$. 
By extending $\q$ to be complex linear, we obtain a Lie
representation $\q_\C$ of the Poisson algebra  $P(M,\C)$ of complex-valued
polynomials on $M$ in $D$.

By assumption the representation of
$\fb$ in $D$ provided by
$\q$ may be integrated to a strongly continuous unitary representation
$\Pi$ of the 1-connected Lie group
$B$ with \la\ $\fb$ in $\h$.
Let $B_\C$ be the universal complexification of
$B$; since $B$ is simply connected, $B_\C$ can be identified with the
1-connected \ss\ 
complex analytic group with \la\ the complexification
$\fb_\C$ of $\fb$. (See Varadarajan [1984], pps. 256--258 and 400--404 for
background on complexifications of Lie groups.)  
Since $B$ is \ss, $B$ is a closed subgroup of $B_\C$, and so we may use 
induction to obtain a strongly continuous unitary representation $\Pi_\C$
of $B_\C$ in a certain \id\ Hilbert space $\K$. 

Now let $C$ be a compact real form of $B_\C$, and denote by $\Gamma$ the
restriction of $\Pi_\C$ to $C$. 
As every strongly continuous unitary representation of a compact
Lie group is completely reducible, we may decompose
\begin{equation*}
\K = {\stackrel{\wedge}{\bigoplus_{i \in I}}}\,\K_i
\label{eq:topds} 
\end{equation*}
for some index set $I \subset \z$, where the \fd\ invariant subspaces
$\K_i$ are the carriers of the irreducible constituents $\Gamma_i$ of
$\Gamma$. Let
$\fc$ be the
\la\ of $C$; then for each $i \in I$, we have the derived
representation $d\Gamma_i$ of $\fc$ in $\K_i$. Set $d\Gamma = \oplus_{i
\in I}\, d\Gamma_i$; this gives a representation of $\fc$ in the
dense subspace
$$D_C = \bigoplus_{i \in I} \K_i.$$

Choose a basis
$\{c_1,\ldots,c_r\}$ of
$\fc$. Since $\fc_\C = \fb_\C$ and as by assumption $\q$ is valued in
$\A$, for every $f \in P(M,\C)$ we may expand
\begin{equation*}
\q_\C(f) = \sum_{n_1,\ldots,n_r} a^f_{n_1 ,\ldots,n_r} \q_\C(c_1)^{
n_1} \cdots \q_\C(c_r)^{ n_r}
\label{eq:polyq} 
\end{equation*}
for some coefficients $a^f_{n_1 ,\ldots,n_r}$.
By means of this formula we can extend the representation
$d\Gamma$ of $\fc$ to a Lie representation $\gamma$
of 
$P(M,\C)$ in $D_C$:
\begin{equation*}
\gamma(f) = \sum_{n_1,\ldots,n_r} a^f_{n_1 ,\ldots,n_r} d\Gamma(c_1)^{
n_1} \cdots d\Gamma(c_r)^{ n_r}
\label{eq:polygamma} 
\end{equation*}
with the \emph{same} coefficients. As each subspace $\K_i$ is invariant,
$\gamma$ restricts to a representation $\gamma_i$ of $P(M,\C)$ in 
$\K_i$. We will show that the existence of these representations
$\gamma_i$ leads to a contradiction.

To this end we recall the following algebraic fact, the proof of which is
given in Gotay, Grabowski, and Grundling [2000].
\begin{lem}
\label{lem:gra} If $L$ is a finite-codimensional Lie ideal of an
infinite-dimen\-sional Poisson algebra $P$ with identity, then either $L$
contains the derived ideal
$\{ P,P\}$ or there is a maximal finite-codimensional 
associative ideal
$J$ of $P$ such that $\{ P,P\}\subset J$. \end{lem}

We apply
Lemma~\ref{lem:gra} to each $L_i = \ker \gamma_i$ which, as $\K_i$ is
finite-dimen\-sional, has finite codimension in $P = P(M,\C)$. First
suppose there is an
$i$ for which 
$\{P,P\} \not\subset L_i$.
Then there must exist a maximal \fcd\ associative ideal $J_i$ in
$P$ with $\{P,P\} \subset J_i.$  If
$\rho$ is the projection $S(\fb_\C) \ra P$, then $I_i =
\rho^{-1}(J_i)$ is a maximal \fcd\ associative ideal in $S(\fb_\C)$ with
$\{S(\fb_\C),S(\fb_\C)\} \subset I_i.$ Since by semisimplicity 
$$\fb_\C =
\{\fb_\C,\fb_\C\} \subset \{S(\fb_\C),S(\fb_\C)\} \subset I_i,$$
and since $1 \not \in I_i$ (as $I_i$ is proper), it follows that  
$I_i$ is the associative ideal generated by $\fb_\C.$ (Actually, this
shows that $S(\fb_\C) = \C \oplus I_i.$)

Since the orbit $M$ is not nilpotent,
there is a nonzero Casimir $\Omega \in S(\fb_\C)$, i.e.
$\rho(\Omega) = \omega$ for some constant $\omega \neq 0.$  Since
$\fb_\C$ is \ss\ it follows from the above observations that $\Omega \in
I_i.$ But then
$\Omega - \omega \not \in I_i$, which is a contradiction since
$\Omega - \omega \in \ker \rho \subset I_i.$

Thus for \emph{every} $i$ it must be the case that $\{P,P\} \subset
L_i$. Again semisimplicity gives
$\fb_\C = \{\fb_\C,\fb_\C\} \subset L_i$, and so 
$\gamma\!\restriction\!\fb_\C = 0$. In particular, then, $d\Gamma = 0$.
Since $\fc$ is a compact real form of $\fb_\C$, the Cartan decomposition
of $\fb_\C$ implies that $d\Pi_\C = 0$. It follows from the induction
construction that the original derived representation
$d\Pi$ of $\fb$ in the domain $D$ must be zero as well. But then
$\q \! \restriction \! \fb = 0$, which contradicts the requirement 
that a quantization represent $\fb$ faithfully.  This concludes the proof
of Theorem~\ref{thm:nogo}.
\endproof

\ms

We remark that Theorem~\ref{thm:nogo} was already known when
$\fb$ is compact (Gotay, Grabowski and Grundling [2000]), in which case
the proof above simplifies greatly and provides an alternate means of
establishing Theorem~2 \emph{ibid.} Notice also that when $\fb$ is compact
every quantization of
$P(M)$ is necessarily polynomial; this follows from the observation that
since $\q \restriction \fb$ is irreducible the representation
space
$\h$ must be
\fd\ together with a well known fact about enveloping algebras
(Prop.~2.6.5 in Dixmier [1976]).


\section{Discussion}

The key observation underlying Theorem~\ref{thm:nogo} is that as $M
\subset
\fb^*$ is non-nilpotent, its ideal $I(M)$ is nonhomogeneous. If $M$
is a basic nilpotent orbit, on the other hand, then $I(M)$ is homogeneous,
and from Theorem 1.1 in Gotay [2002] we know that there do exist
polynomial quantizations of
$P(M)$. (Although it is not clear to what extent these are ``nontrivial''
in general.) Taken together, these two results serve to establish a
conjecture of Gotay [2000] when
$\fb$ is
\ss :
\emph{There exists a consistent polynomial quantization of $P(M)$ if and
only if $I(M)$ is homogeneous.}

Finally, we remark that the restriction to basic coadjoint orbits is
for physical reasons, cf. Gotay [2000]. For our
purposes here, we may delete the adjective ``basic'' and
consider arbitrary orbits. Both Theorem~\ref{thm:nogo} and the conjecture
above will remain valid in this extended context.


\section*{References}

\begin{description}

\item Dixmier, J. [1977] {\sl Enveloping Algebras.} (North Holland,
Amsterdam).

\item Gotay, M.~J. [2000] Obstructions to quantization.
In: 		 {\sl Mechanics: {}From Theory to Computation (Essays in Honor
of Juan-Carlos Simo)}, J. Nonlinear Sci. Eds.  (Springer, New York),
171--216.

\item Gotay, M.~J. [2002] On quantizing semisimple basic algebras, I:
sl$(2,\r)$. To appear in: 		 {\sl The Jerry Marsden 60$^{th}$ Birthday
Volume}, P. Holmes, P. Newton, and A. Weinstein, Eds.  (Springer, New
York). Preprint math-ph/0012034.

\item Gotay, M.~J., Grabowski, J., and Grundling, H.~B.
[2000] An obstruction to quantizing compact symplectic manifolds. {\it
Proc. Amer. Math. Soc.} {\bf 28}, 237--243.

\item Varadarajan, V.~S. [1984] {\sl Lie Groups, Lie Algebras, and their
Representations.} (Springer, New York).

\end{description}


\end{document}